\documentclass[journal]{IEEEtran}
\IEEEoverridecommandlockouts
\usepackage{mathrsfs}
\usepackage{amsfonts}
\usepackage{ifpdf}
\usepackage{cite}

\ifCLASSINFOpdf
\usepackage[pdftex]{graphicx}
\else \fi
\usepackage[cmex10]{amsmath}
\usepackage{array}
\usepackage{mdwmath}
\usepackage{mdwtab}
\usepackage{eqparbox}
\usepackage[tight,footnotesize]{subfigure}
\usepackage{algorithmic}
\usepackage{algorithm}
\usepackage{amsmath}
\usepackage{epsfig}
\usepackage{ae}
\usepackage{amssymb}
\usepackage{times}
\usepackage{amsmath}   
\usepackage{amsfonts} 
\usepackage{rotating} 
\usepackage{mathrsfs} 
\usepackage{xcolor}

\usepackage{amsthm} 

\usepackage{supertabular} 

\usepackage{multirow}
\newlength\savewidth

\newtheorem{theorem}{\textbf{Theorem}}
\newtheorem{lemma}{\textbf{Lemma}}

\newtheorem{definition}{\textbf{Definition}}
\newtheorem{remark}{\textbf{Remark}}

\begin{document}

\title{To Relay or not to Relay: Open Distance and Optimal Deployment for Linear Underwater Acoustic Networks}
\author{
         Yuzhou~Li,~\IEEEmembership{Member,~IEEE},
         Yu~Zhang,
         Hongkuan~Zhou,
         and~Tao~Jiang,~\IEEEmembership{Senior Member,~IEEE}
\thanks{This work was supported in part by the National Science Foundation of China with Grants 61601192, 61601193, 61729101, and 61631015, the Major Program of National Natural Science Foundation of Hubei in China with Grant 2016CFA009, and the Fundamental Research Funds for the Central Universities with Grant 2016YXMS298.}
\thanks{The authors are with the School of Electronic Information and Communications, Huazhong University of Science and Technology, Wuhan, 430074, P. R. China (e-mail: \{yuzhouli, yu\_zhang, hongkuanzhou, taojiang\}@hust.edu.cn).}
}
\maketitle
\IEEEpeerreviewmaketitle
\begin{abstract}
Existing works have widely studied relay-aided underwater acoustic networks under some specialized relay distributions, e.g., equidistant and rectangular-grid. In this paper, we investigate two fundamental problems that under which conditions a relay should be deployed and where to deploy it if necessary in terms of the energy and delay performance in linear underwater acoustic networks. To address these two problems, we first accurately approximate the complicated effective bandwidth and transmit power in the logarithm domain to formulate an energy minimization problem. By analyzing the formulation, we discover a critical transmission distance, defined as open distance, and explicitly show that a relay should not be deployed if the transmission distance is less than the open distance and should be otherwise. Most importantly, we derive a closed-form and easy-to-calculate expression for the open distance and also strictly prove that the optimal placing position is at the middle point of the link when a relay should be introduced. Moreover, although this paper considers a linear two-hop relay network as the first step, our derived results can be applied to construct energy-efficient and delay-friendly multi-hop networks. Simulation results validate our theoretical analysis and show that properly introducing a relay can dramatically reduce the network energy consumption almost without increasing the end-to-end delay.
\end{abstract}

\section{Introduction}
As the only effective means for long-range data delivering in extremely hostile undersea environments, acoustic communications have been widely used in various underwater devices and scenarios. However, it is well-known that acoustic channels are quite bandwidth-limited and underwater acoustic transmission is usually energy-expensive, especially in long-range transmission cases \cite{UOWC_ASurvey_ieeeCST2017,UnderwaterSensorNetwork_HDai_globecom2015}. To meet the ever-increasingly high-data-rate and energy-efficient (and thus lifetime-enhanced) demands imposed by real-time or long-term marine applications such as marine rescues and monitoring, a potential solution is to introduce relays between transmitters and receivers to construct multi-hop networks. This is because deploying relays to shorten hop distance can expand the effective bandwidth and reduce the transmit power of each hop \cite{UnderwaterAcousticCommunication_CapacityVSDistance_ACMSIGMOBILE2007,AdmissionControlResourceAllocation_D2DNetworks_JSAC2016,UnderwaterAcousticCommmunication_ARelayScenarios_Oceans2007}. Nevertheless, relay deployment is a double-edged sword, although having the aforementioned benefits, it may in turn degrade the network performance as the deployed relays will spend extra time on packet forwarding and consume additional energy to receive data \cite{DMtradeoff_ieeeTC2009,CooperativeCommun_ieeeWC2010,FullDuplexRelaying_ieeeTCom2016,DoFRelayChannel_ieeeCOM2017}. It is thus important to understand how relay deployment affects the performance of underwater acoustic networks and how to quantify these impacts.

There have been extensive works to investigate relay deployment problems in underwater acoustic networks from different perspectives, e.g., optimizing the relay number \cite{UnderwaterAcousticCommmunication_ARelayScenarios_Oceans2007,UnderwaterAcousticNetwork_OptimalRelayNumber_ieeeJOE2010,
UnderwaterAcousticCommunication_optimizingRelayNumbers_ieeeSJ2016,UnderwaterAcousticSensorNetworks_DeploymentAnalysis_ACM2006} and determining the best relay position \cite{UnderwaterAcousticCommunication_RelayCapacity_ieeeWCNC2010,UnderwaterAcousticNetworks_EnergyEfficientRouting_ieeeJSAC2008,
UnderwaterSensorNetworks_NodeReplacement_ieeeJOE2013,UnderwaterAcousticNetworks_relayPlacement_ieeeJOE2014}. In \cite{UnderwaterAcousticSensorNetworks_DeploymentAnalysis_ACM2006}, Pompili \textit{et al.} studied the problem of how many sensors at least are required to achieve the optimal sensing and communication coverage in two- and three-dimensional underwater acoustic sensor networks. By minimizing an elaborated cost function,
analytical solutions for the optimal number of relays were derived in \cite{UnderwaterAcousticCommmunication_ARelayScenarios_Oceans2007,UnderwaterAcousticNetwork_OptimalRelayNumber_ieeeJOE2010,
UnderwaterAcousticCommunication_optimizingRelayNumbers_ieeeSJ2016} to evaluate the tradeoffs involved in the design of a linear relay acoustic link. In \cite{UnderwaterAcousticCommunication_RelayCapacity_ieeeWCNC2010}, Cao \textit{et al.} obtained that the relay location is a much more critical factor with respect to the system capacity than power allocation from simulation results. Leveraging an optimal per-hop distance from numerical observations, Zorzi \textit{et al.} \cite{UnderwaterAcousticNetworks_EnergyEfficientRouting_ieeeJSAC2008} developed a relay selection routing algorithm to minimize the system's total energy consumption. Routing scheme and relay replacement were jointly optimized in \cite{UnderwaterSensorNetworks_NodeReplacement_ieeeJOE2013} to improve the energy efficiency of underwater acoustic sensors networks for a given rectangular-grid node distribution.
Further, Kam \textit{et al.} \cite{UnderwaterAcousticNetworks_relayPlacement_ieeeJOE2014} considered a random node distribution scenario and proved that equal spacing from the source to the destination is optimal when using a defined globally optimal frequency.

To summarize, \cite{UnderwaterAcousticCommmunication_ARelayScenarios_Oceans2007,UnderwaterAcousticCommunication_RelayCapacity_ieeeWCNC2010,
UnderwaterAcousticNetwork_OptimalRelayNumber_ieeeJOE2010,UnderwaterAcousticCommunication_optimizingRelayNumbers_ieeeSJ2016,
UnderwaterAcousticNetworks_EnergyEfficientRouting_ieeeJSAC2008,UnderwaterSensorNetworks_NodeReplacement_ieeeJOE2013,
UnderwaterAcousticSensorNetworks_DeploymentAnalysis_ACM2006} investigated relay-related performance under some specialized relay distributions, e.g., equidistant and rectangular-grid, but there are not analyses or proofs as to why such distribution is optimal. Although \cite{UnderwaterAcousticNetworks_relayPlacement_ieeeJOE2014} strictly proved that equidistant spacing is optimal for energy efficiency in randomly-placed relay networks, its results hold only for narrow-band signals with bandwidth of 0.1 kHz. Besides, a deceptively-simple but fundamental problem, whether it is necessary to deploy relays in underwater acoustic transmission, still remains unsolved and few works have discussed this problem to the best of our knowledge. If the answer is necessary, a concomitant fundamental problem is what is the optimal relay deployment, e.g., equidistant in \cite{UnderwaterAcousticCommunication_RelayCapacity_ieeeWCNC2010,
UnderwaterAcousticNetwork_OptimalRelayNumber_ieeeJOE2010,UnderwaterAcousticCommunication_optimizingRelayNumbers_ieeeSJ2016,UnderwaterAcousticSensorNetworks_DeploymentAnalysis_ACM2006,
UnderwaterAcousticNetworks_EnergyEfficientRouting_ieeeJSAC2008}, in terms of the selected performance indexes. In view of these, this paper devotes to quantitatively addressing these two fundamental problems, namely under which conditions relays should be deployed and where to deploy them if necessary. Assuming that each hop delivers data over its full effective bandwidth, we answer them in terms of the energy consumption and end-to-end delay by considering direct and linear two-hop relay transmission scenarios as the first step. By analyzing our formulated energy minimization problem, we theoretically derive a critical transmission distance with a closed-form expression, defined as the open distance in this paper, which provides an extremely simple way to fast judge whether to deploy a relay. In the case when a relay is needed, we further obtain the optimal relay placing position.

The main contributions of this work are as follows:
\begin{itemize}
\item We first accurately approximate the very complicated effective bandwidth and transmit power in the logarithm domain to formulate an energy minimization problem, instead of adopting traditionally numerical evaluation approaches (by which quantitative results can hardly be derived), for comparing the energy consumption and delay performance between direct and linear two-hop relay transmission schemes.
\item Based on the formulation, we discover that there exists a critical transmission distance, defined as open distance, which provides an extremely simple method to decide whether a relay should be deployed. Specifically, a relay should not be deployed when the transmission distance is less than the open distance and should be otherwise.
\item Most importantly, we leverage the differential analysis to derive a closed-form and easy-to-calculate expression for the open distance and also strictly prove that the optimal placing position is at the middle point of the link when a relay should be introduced.
\item Although this paper considers a linear two-hop relay network as the first step, our derived results provide significant guidelines for constructing energy-efficient and delay-friendly multi-hop networks. Specifically, relays are needed to be deployed at the midpoint of each hop for saving energy and maintaining the end-to-end delay until the length between any two adjacent nodes does not exceed the open distance.
\item Extensive simulation results validate our theoretical analysis and show that properly introducing a relay can dramatically reduce the network energy consumption (up to 71.77\%) almost without increasing the end-to-end delay (less than 1.56\%). Furthermore, we apply a polynomial fitting method to derive another precise expression for the open distance through least-squares approximation based on the sufficient realistic data for potential applications.
\end{itemize}

The remainder of this paper is organized as follows. In Section~\ref{Section:BandPT}, we overview the basic knowledge about underwater acoustic channels.
In Section~\ref{Section:SystemModel}, we introduce system scenarios, formulate the concerned problem, and present quantitative results.
Section~\ref{Section:Solution} provides the solution proof and analysis and extensive simulation results are presented in Section~\ref{Section:Numeric Results}.
Finally, we conclude our paper in Section~\ref{Sec:conclusion}.

\section{Effective Bandwidth and Transmit Power} \label{Section:BandPT}
In this section, we first introduce the path loss and the ambient noise of an underwater acoustic channel, based on which we obtain the effective bandwidth and required transmit power for a given signal-to-noise ratio (SNR). We then precisely approximate them  for more easily characterizing the bandwidth-range and power-range dependent features of the underwater acoustic channel.
\subsection{Path Loss} \label{Subsection:PathLoss}
According to \cite{UnderwaterSound_principles_1967}, path loss of an underwater acoustic channel over a distance $l$ in km for a signal at frequency $f$ in kHz, denoted by dimensionless $A(l,f)$, can be modeled in dB form as
\begin{equation}\label{Eq:PathLoss}
  10\log_{10}A(l,f) = k \cdot 10\log_{10}(l\cdot10^3) + l \cdot 10\log_{10}a(f)
\end{equation}
where $k$ is the spreading factor that defines the geometry of the acoustic propagation, commonly $k=1$ for cylindrical spreading, $k=2$ for spherical spreading, and $k=1.5$ for practical spreading. In addition, $10\log_{10}a(f)$ represents the absorption coefficient in dB/km, which, from \cite{UnderwaterAcoustic_fundamentals_JASA1991}, is expressed by the Thorp's formula as
\begin{equation}\label{Eq:AbsorptionCoefficient}
  10\log_{10}a(f) = \frac{0.11f^2}{1+f^2} + \frac{44f^2}{4100+f^2} + \frac{2.75f^2}{10^{4}} + 0.003.
\end{equation}

In (\ref{Eq:PathLoss}), the first term on the right side denotes the spreading attenuation that increases with the distance $l$, and the second term represents the absorption loss related to both the distance $l$ and frequency $f$. As a result, path loss of an underwater acoustic channel is not only dependent on the transmission distance, but also on the signal frequency. This feature distinguishes underwater acoustic from terrestrial radio transmission as the radio suffers negligible absorption loss in the air.

\subsection{Ambient Noise} \label{Subsection:AmbientNoise}
Ambient noise exists all the time in the ocean, mainly including four parts of turbulence, shipping, waves, and thermal noise and usually showing Gaussian characteristics.
From \cite{UnderwaterAcoustic_fundamentals_JASA1991}, empirically continuous power spectral density (p.s.d.) models of these four parts in $\mu \ \text Pa$ per Hz at a signal frequency $f$ in kHz, denoted by $N_{t}(f)$, $N_{s}(f)$, $N_{w}(f)$, and $N_{\text{th}}(f)$, respectively, are given by
\begin{equation}\label{Eq:NoiseComponent}
\begin{aligned}
10\log_{10}N_{t}(f)  \ \ =\ \ &17 - 30\log_{10}f\\
10\log_{10}N_{s}(f)  \ \ =\ \ &40 + 20(s-0.5) + 26\log_{10}f -\\
                      & 60\log_{10}(f+0.03)\\
10\log_{10}N_{w}(f)  \ \ =\ \ &50 + 7.5w^{1/2} + 20\log_{10}f -\\
                      & 40\log_{10}(f+0.4)\\
10\log_{10}N_{\text{th}}(f) \ \ =\ \ &-15 + 20\log_{10}f
\end{aligned}
\end{equation}
where $s$ is the shipping activity factor between 0 for low activity and 1 for high activity, and $w$ is the wind speed in the unit of m/s.

Adding the four components in (\ref{Eq:NoiseComponent}) together, we get the p.s.d. of the overall ambient noise of the system in $\mu \ \text Pa$ per Hz as
\begin{equation}\label{Eq:AmbientNoise}
  N(f) = N_{t}(f) + N_{s}(f) + N_{w}(f) + N_{\text{th}}(f).
\end{equation}
Numerical results\footnote{Throughout this paper, numerical results or numerical values are obtained by directly using (\ref{Eq:PathLoss})--(\ref{Pt}) without any simplification or approximation.} in \cite{UnderwaterAcousticCommunication_CapacityVSDistance_ACMSIGMOBILE2007} show that the turbulence noise produces influences only in the frequency region of $f<10$ Hz, the noise from distant shipping and wind-driven waves becomes dominant when $f$ lies in 10 Hz--100 Hz and 100 Hz--100 kHz, respectively, and the thermal noise holds the major proportion when $f>100$ kHz.

\begin{figure}[t]
\centering \leavevmode \epsfxsize=3.5 in  \epsfbox{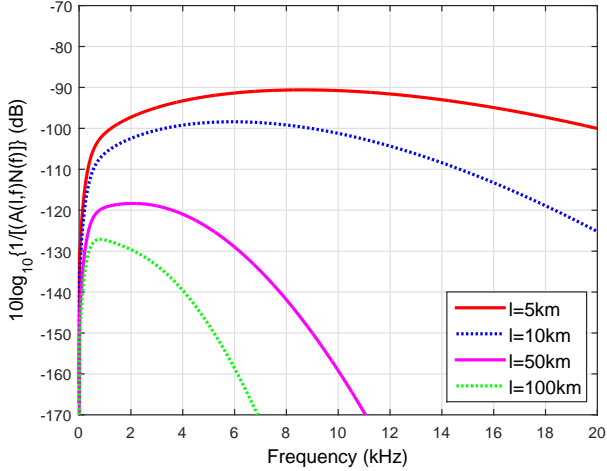}
\centering \caption{Frequency-dependent feature of the narrow-band SNR for different transmission distances, denoted by $10\log_{10}\{1/[(A(l,f)N(f)]\}$ in dB form. In this figure, moderate shipping activity ($s=0.5$) and no wind ($w=0$) are applied in the noise p.s.d. (\ref{Eq:AmbientNoise}), and practical spreading ($k=1.5$) is used for the path loss model (\ref{Eq:PathLoss}).} \label{Fig:ANProduct}
\end{figure}

\begin{figure}[t]
\centering \leavevmode \epsfxsize=3.5 in  \epsfbox{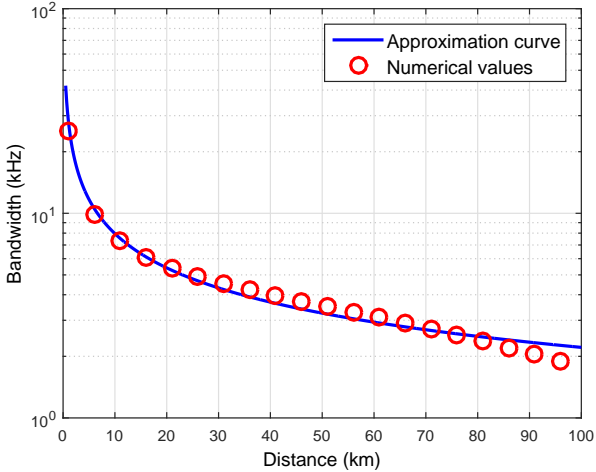}
\centering \caption{Comparison of the effective bandwidth between approximate results from (\ref{Eq:PT}) and numerical values, where $\lambda=0.5392$ and $\omega=10^{1.4291}$.} \label{Fig:Bandwidth}
\end{figure}

\subsection{Effective Bandwidth and Transmit Power} \label{Subsection:rangeFeature}
\begin{figure}[t]
\centering \leavevmode \epsfxsize=3.5 in  \epsfbox{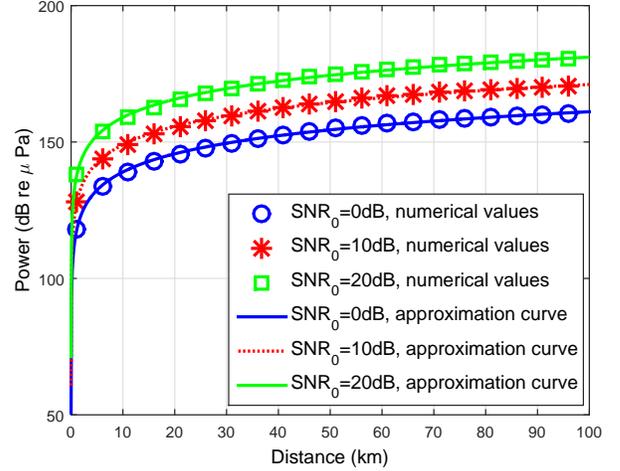}
\centering \caption{Comparison of the required transmit power under different target SNRs, between approximate results from (\ref{Eq:uPaToWatt}) and numerical values, where $\gamma=2.2074$ for all target SNRs and $\psi=10^{0.1\rm{SNR_0}-4.9040}$.} \label{Fig:PT}
\end{figure}

\begin{figure}[t]
\centering \leavevmode \epsfxsize=3.5 in  \epsfbox{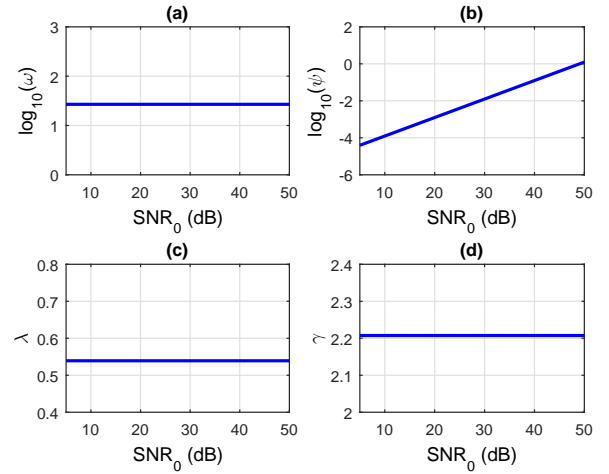}
\centering \caption{Model parameters vs. target SNRs. (a) $\log_{10}(\omega)$ vs. $\rm{SNR_0}$. (b) $\log_{10}(\psi)$ vs. $\rm{SNR_0}$. In this subfigure, the curve can be further approximated as a line with a slope of $0.1000$ and a vertical intersection of $-4.9040$. (c) $\lambda$ vs. $\rm{SNR_0}$. (d) $\gamma$ vs. $\rm{SNR_0}$. For all the subfigures, moderate shipping activity ($s=0.5$) and no wind ($w=0$) are applied in the noise p.s.d. (\ref{Eq:AmbientNoise}), and practical spreading ($k=1.5$) is used for the path loss model (\ref{Eq:PathLoss}).} \label{Fig:ModelParameters}
\end{figure}

Using the path loss model (\ref{Eq:PathLoss}) and ambient noise model (\ref{Eq:NoiseComponent}), the narrow-band SNR, which is a dimensionless measure, is given by
\begin{equation}\label{Eq:NarrowSNR}
  {\rm SNR}(l,f) = \frac{S_l(f)\Delta f/A(l,f)}{N(f)\Delta f} = \frac{S_l(f)}{A(l,f)N(f)}
\end{equation}
where $S_l(f)$ is the p.s.d. of the transmit signal and $\Delta f$ is a narrow frequency band around $f$ \cite{UnderwaterAcousticCommunication_CapacityVSDistance_ACMSIGMOBILE2007,UnderwaterAcousticCommmunication_ARelayScenarios_Oceans2007}. The denominator $A(l,f)N(f)$, illustrated in Fig.~\ref{Fig:ANProduct}, determines the frequency-dependent feature of the narrow-band SNR.
From Fig.~\ref{Fig:ANProduct}, there exists an optimal frequency $f_{0}(l)$ for a given transmission distance $l$, where $1/[(A(l,f)N(f)]$ (and thus ${\rm SNR}(l,f)$ for a given $S_l(f)$) reaches its maximum. Around the maximum of $1/[(A(l,f)N(f)]$, we adopt the 3-dB bandwidth as the effective bandwidth of the transmission denoted by $B_{3}(l)$, which is the frequency range that satisfies $\{1/[A(l,f)N(f)]\}>\{1/[2A(l,f_{0}(l))N(f_{0}(l))]\}$. Suppose the transmitter works on the whole 3-dB bandwidth, then the overall SNR at the receiver is given by
\begin{equation}\label{Eq:SNR}
  {\rm SNR}(l,B_{3}(l)) = \frac{\int_{B_{3}(l)}S_{l}(f)A^{-1}(l,f)df}{\int_{B_{3}(l)}N(f)df}.
\end{equation}
Assume that the power of the transmit signal, denoted by $P_{t}(l)$ in $\mu$ Pa, is  equally distributed over the 3-dB bandwidth, the required transmit power to satisfy a target SNR $\rm SNR_{0}$ at a distance $l$ in km is given by
\begin{equation}\label{Pt}
P_{t}(l) = 10^{3}\cdot B(l){\rm SNR_{0}}\frac{\int_{B_{3}(l)}N(f)df}{\int_{B_{3}(l)}A^{-1}(l,f)df}
\end{equation}
where $B(l)$ in kHz is the size of frequency range $B_{3}(l)$.

From the numerical evaluations of \cite{UnderwaterAcousticCommunication_CapacityVSDistance_ACMSIGMOBILE2007} and \cite{UnderwaterAcousticCommmunication_ARelayScenarios_Oceans2007}, both the effective bandwidth $B(l)$ and the required transmit power $P_{t}(l)$ can be precisely approximated as functions of the transmission distance $l$ for a given SNR as follows
\begin{equation}\label{Eq:PT}
\begin{aligned}
  B(l) = \omega l^{-\lambda} \\
  P_{t}(l) =  \delta l^{\gamma}
\end{aligned}
\end{equation}
where the units of $B(l)$, $P_{t}(l)$, and $l$ are kHz, $\mu$ Pa, and km, respectively. Furthermore, from \cite{UnderwaterSound_principles_1967}, the conversion of the transmit power $P_{t}(l)$ in $\mu$ Pa to its corresponding electrical power $P_{T}(l)$ in Watt is given by
\begin{equation}\label{Eq:uPaToWatt}
  P_{T}(l) = P_{t}(l) \cdot (10^{-17.2}/\eta) = \psi \cdot l^{\gamma}
\end{equation}
where $\rm{10^{-17.2}}$ is the conversion factor and $\eta$ denotes the overall efficiency of the power amplifier and transducer.
In (\ref{Eq:PT}) and (\ref{Eq:uPaToWatt}), all the model parameters, including the scaling factors $\omega$ and $\psi$ as well as the exponential coefficients $\lambda$ and $\gamma$, are positive values and can be readily obtained by first-order least-squares polynomial approximation on a logarithmic scale. Following this principle, we fit $B(l)$ and $P_{T}(l)$ under different target SNRs in Fig.~\ref{Fig:Bandwidth} and Fig.~\ref{Fig:PT}, respectively, from which excellent matches can be reached by (\ref{Eq:PT}) and (\ref{Eq:uPaToWatt}).

To further evaluate how the model parameters depend on the target SNR and determine their values, we apply the same approximation method for all those target SNRs of interests and plot their variation tendency with the target SNRs in Fig.~\ref{Fig:ModelParameters}. From Fig.~\ref{Fig:ModelParameters}, except that $\psi$ increases linearly as the target SNR grows, the other parameters are all invariant to the target SNRs. More specifically, the numerical scales of $\omega$, $\psi$, $\lambda$, and $\gamma$, denoted by sets $\Omega$, $\Psi$, $\Lambda$, and $\Gamma$, respectively, lie in the following regions
\begin{equation}\label{Eq:parameterScale}
\begin{aligned}
  \Omega &: \omega > 0 \\
  \Psi &: \psi = 10^{0.1\rm{SNR_0}-4.9040} \\
  \Lambda &: 0.5 < \lambda < 0.6 \\
  \Gamma &: 2.1 < \gamma < 2.3.
\end{aligned}
\end{equation}
Eq.~(\ref{Eq:parameterScale}) is important to solve the optimization problem (\ref{Eq:Problem_Formulation}) and will be used throughout the entire derivation process in Section~\ref{Section:Solution} to identify the sign of some important expressions.

At this point, we have reviewed the channel properties of underwater acoustic propagation including the path loss and the ambient noise. From Figs.~\ref{Fig:ANProduct}, \ref{Fig:Bandwidth}, and \ref{Fig:PT}, we can further obtain the following two important features regarding the underwater acoustic channels:
\begin{itemize}
\item \textbf{Bandwidth-range dependent feature}. As shown in Fig.~\ref{Fig:Bandwidth}, the effective bandwidth of underwater acoustic channels decreases exponentially when the transmission distance grows. For example, the effective bandwidth can reach dozens of kHz when the transmission distance is 1 km, while it becomes less than 10 kHz if the distance exceeds 10 km.
\item \textbf{Power-range dependent feature}. Opposite to the bandwidth, the required transmit power of an underwater acoustic channel to satisfy a given SNR increases exponentially when the transmission distance grows, as depicted in Fig.~\ref{Fig:PT}. For instance, when communicating with a target 1 km away at the $\rm{SNR_0}$ of 20 dB, the required transmit power is less than 1 W, but it surges to about 100 W when the distance is 100 km.
\end{itemize}

These facts, i.e., the bandwidth-range and power-range dependent features of underwater acoustic channels, imply that the transmission performance possibly can be improved by deploying relays along an underwater acoustic link. This is because deploying relays can significantly expand the effective bandwidth and reduce the transmit power through shortening hop distance (see Figs.~\ref{Fig:Bandwidth} and \ref{Fig:PT}), and thus may increase the data rate and decrease the network energy consumption in the meantime. However, relay deployment in turn may degrade the network performance as relays will spend extra time on packet forwarding and consume additional energy to receive data. It is thus important to understand under which conditions relays should be introduced and where they are deployed if necessary, which will be quantitatively discussed in Section~\ref{Section:SystemModel}.

\section{System Scenarios, Problem Formulation, and Quantitative Results} \label{Section:SystemModel}
In this section, we introduce the considered system scenarios and involved parameters, formulate an energy minimization problem to address the above two fundamental problems, and present quantitative results to answer them.
\subsection{Description of Linear Underwater Acoustic Networks} \label{Subsection:SingleCellScenarios}
\begin{figure}[t]
\centering \leavevmode \epsfxsize=3.5 in  \epsfbox{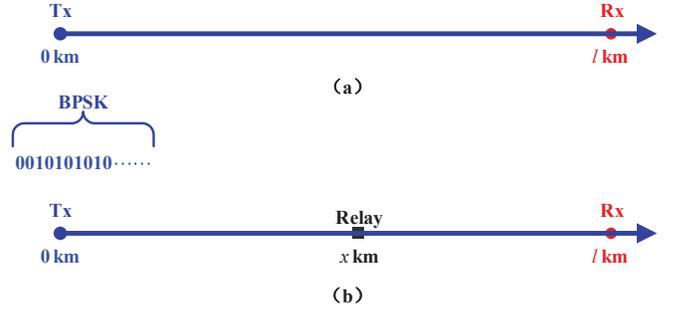}
\centering \caption{System scenario. (a) Direct acoustic transmission from the Tx to the Rx. (b) Linear Two-hop relay acoustic transmission with one relay located at $x$ km on the line between the Tx and the Rx.} \label{Fig:Scenario}
\end{figure}

Consider an underwater acoustic link, as shown in Fig.~\ref{Fig:Scenario}(a), where a packet of $L$ bits needs to be delivered from the source to the destination, denoted by Tx and Rx and located at the origin and the position $l$ km ($l>0$) away on the axis, respectively. In this scenario, we investigate under which conditions a relay should be deployed on the line between the Tx and the Rx and where it should be deployed (i.e., determine the value of $x$, see Fig.~\ref{Fig:Scenario}(b)) in terms of the overall energy consumption and the end-to-end delay\footnote{It is worthwhile to note that the analytical results for single-relay deployment problems are also significant for determining the best deploying schemes in multiple-relay cases, which are extremely complicated and remains unsolved at present. In particular, Section~\ref{Subsection:Summary} will discuss how to apply our derived results to construct energy-efficient multi-hop networks.}. We refer this kind of architectures as linear underwater acoustic networks, namely either the Tx, the relay, or the Rx is on a straight line.

Specifically, the end-to-end delay is defined as the sum of the duration (i.e., consumed time) of radiating amount of bits from the buffer and that of propagating over a distance \cite{EE_Delay_D2D_JSAC2013,Through_Delay_EE_TWC2014}. Unlike terrestrial radio traveling at the speed of light, sound propagates underwater at a very low speed of about 1500 m/s. As a result, the end-to-end delay of underwater acoustic communications is dominated by the propagating delay, another feature that distinguishes underwater acoustic from terrestrial radio transmission.
In the case of the direct transmission (See Fig.~5(a)), the end-to-end delay to deliver a packet of $L$ bits for a given SNR over a distance of $l$ km can be calculated as
\begin{equation}\label{Eq:DirectDelay}
  t_{0}(l) = t_{\text{Radiate}} + t_{\text{Propagate}} = \frac{L}{\alpha B(l)} + \frac{l}{c}
\end{equation}
where $c$ is the acoustic speed, about 1500 m/s, and $\alpha$ is the bandwidth efficiency of the modulation in bps/Hz (e.g., $\alpha = 1$ bps/Hz for BPSK and $\alpha = 2$ bps/Hz for QPSK). Accordingly, the overall energy consumption in Joule, defined as the sum of the transmit and receive energy, is modeled as
\begin{equation}\label{Eq:DirectEnergy}
\begin{aligned}
  E_{0}(l) &= E_{\text{Transmit}} + E_{\text{Receive}} \\
  &\overset{1}{=} P_{T}(l)\frac{L}{\alpha B(l)} + P_{R}\frac{L}{\alpha B(l)} \\
          &\overset{2}{=} \frac{L}{\alpha \omega} \cdot \big[\psi l^{\lambda+\gamma} + P_{R}l^{\lambda}\big].
\end{aligned}
\end{equation}
where $\overset{2}{=}$ is obtained by substituting (\ref{Eq:PT}) and (\ref{Eq:uPaToWatt}) into $\overset{1}{=}$, and $P_{R}$ is a distance-independent constant parameter to denote the receiver power \cite{UnderwaterNerwoks_PracticalIssues_ACM2007}.

In the case of two-hop relay transmission (see Fig.~\ref{Fig:Scenario}(b)), the end-to-end delay and the overall energy consumption over the two-hop path can be computed by adding those incurred by the two individual hops. Specifically, the end-to-end delay is given by
\begin{equation}\label{Eq:RelayNetworksDelay}
  t_{1}(l) = t_{0}(x) + t_{0}(l-x) = \frac{L}{\alpha B(x)} + \frac{L}{\alpha B(l-x)} + \frac{l}{c}
\end{equation}
and the overall energy consumption is\footnote{Eq.~(\ref{Eq:RelayNetworksDelay}) and Eq.~(\ref{Eq:RelayNetworksEnergy}) indicate that the transmission bandwidth and the transmit power can be adjusted arbitrarily according to the transmission distance. This may be infeasible in practice but the corresponding analysis is theoretically significant as it can provide a performance bound to what can be achieved and thus has been widely adopted in existing works, e.g., in \cite{UnderwaterAcousticCommmunication_ARelayScenarios_Oceans2007,UnderwaterAcousticCommunication_RelayCapacity_ieeeWCNC2010,
UnderwaterAcousticNetwork_OptimalRelayNumber_ieeeJOE2010,UnderwaterAcousticCommunication_optimizingRelayNumbers_ieeeSJ2016,
UnderwaterAcousticNetworks_EnergyEfficientRouting_ieeeJSAC2008,UnderwaterSensorNetworks_NodeReplacement_ieeeJOE2013,
UnderwaterAcousticSensorNetworks_DeploymentAnalysis_ACM2006}.}
\begin{equation}\label{Eq:RelayNetworksEnergy}
\begin{aligned}
  &E_{1}(x) = E_{0}(x) + E_{0}(l-x) \\
  &= \frac{L[P_{T}(x)+P_{R}]}{\alpha B(x)} + \frac{L[P_{T}(l-x)+P_{R}]}{\alpha B(l-x)}\\
  &=  \frac{L}{\alpha \omega} \cdot \big[\psi x^{\lambda+\gamma} + P_{R}x^{\lambda} + \psi(l-x)^{\lambda+\gamma} + P_{R}(l-x)^{\lambda}\big].
\end{aligned}
\end{equation}

From (\ref{Eq:RelayNetworksEnergy}), we can obtain the following interesting properties with respect to $E_{1}(x)$, which are totally in accordance with our intuitional understanding.
\begin{itemize}
\item \textbf{Positivity:} $E_{1}(x) > 0$ holds for all $x\in (0,l)$. This directly follows the fact that delivering data implies energy consumption.
\item \textbf{Symmetry:} $E_{1}(x)$ is symmetric with respect to $x = l/2$ as $E_{1}(x) = E_{1}(l-x)$, $ \forall x\in (0,l)$. It thus suffices to consider half of the total distance when investigating the relay-related performance.
\item \textbf{Inclusivity:} Direct transmission can be seen as a special case of two-hop relay transmission by locating the relay at the source or the destination\footnote{
Note that the $n$-order derivatives of $E_{1}(x)$ may be meaningless at $x = 0$ and $x = l$. For rigorous utilization of the differential analysis in Section~\ref{Section:Solution}, we also use $x = 0^+$ and $x = l^-$ to equivalently represent the locations at the source and the destination, respectively.}, because
      \begin{equation}\label{Eq:limitValue}
      \begin{aligned}
      \lim\limits_{x\rightarrow{0^+}}E_{1}(x) &= \lim\limits_{x\rightarrow{l^-}}E_{1}(x)
        = \frac{L}{\alpha \omega}\big\{\psi l^{\lambda+\gamma} + P_{R}l^{\lambda}\big\} \\
        &= E_{1}(0) = E_{1}(l) = E_{0}(l).
      \end{aligned}
      \end{equation}
\end{itemize}

\subsection{Problem Formulation}\label{Subsection:ProblemFormulation}
From (\ref{Eq:DirectDelay}) and (\ref{Eq:DirectEnergy}), $t_{0}(l)$ and $E_{0}(l)$ can be easily calculated. To determine which kind of transmission (direct or two-hop relay) is the best, it is thus needed to know the minimum energy consumption and end-to-end delay of the two-hop relay transmission. Regarding the delay performance, we will discuss it by simulation results in Section~\ref{Section:Numeric Results}. From the energy consumption perspective, mathematically, the following two questions are required to be quantitatively answered.
\begin{enumerate}[]
\item \textbf{\textit{Question 1: When should a relay  be introduced?}} A relay should not be deployed if $E_{0}(l) \leq E_{1}(x)$ for all $x\in (0,l)$. Otherwise, there at least exists an $ x' \in (0,l)$ such that $E_{0}(l) >  E_{1}(x')$. In other words, a proper relay deployment will consume less energy than the direct transmission.
\item \textbf{\textit{Question 2: Where should a relay be deployed if necessary?}} A relay should be deployed at the position $x_0$ where the energy cost reaches its minimum, i.e.,  $E_{1}(x_{0}) = \min_x E_{1}(x)$ or $x_{0} = \arg \min_x E_{1}(x)$.
\end{enumerate}

Hence, we need to find out the minimum value of $E_{1}(x)$ and compare it with $E_{0}(l)$ to decide whether a relay is necessary, which can be obtained by solving the following energy minimization problem
\begin{equation}\label{Eq:Problem_Formulation}
\begin{aligned}
{\min_x }       ~~&E_{1}(x) \\
\text{s.t.}   ~~&\text{C1:} ~~0 \le x \le l
\end{aligned}
\end{equation}
In (\ref{Eq:Problem_Formulation}), all involved parameters are in their individual ranges. Specifically, $\gamma$, $\omega$, $\lambda$, and $\psi$ used in the fitting models of bandwidth $B(l)$ and transmit power $P_{T}(l)$ (see (\ref{Eq:PT}) and (\ref{Eq:uPaToWatt})) are determined by (\ref{Eq:parameterScale}), the receive power $P_{R} > 0$, the bandwidth efficiency $\alpha > 0$, and the packet size $L \in \mathbb{N^{+}}$.

\subsection{Open Distance}\label{Subsection:openDistance}

\begin{figure}[t]
\centering \leavevmode \epsfxsize=3.5 in  \epsfbox{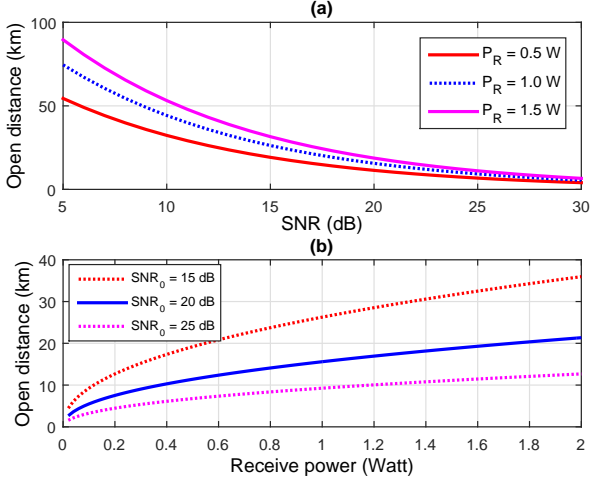}
\centering \caption{(a) Open distance vs. target SNRs for given receive powers. (b) Open distance vs. receive power for given target SNRs. In the figure, we set $\lambda =0.5392$ and $\gamma = 2.2074$.} \label{Fig:TheoOpenDistance}
\end{figure}

To quantify the optimal solution of the problem (\ref{Eq:Problem_Formulation}), we first define a concept of open distance.
\begin{definition} \label{Definition:OpenDistance}
An open distance, denoted by $l_{\text{OP}}$, determines a largest transmission distance between the Tx and the Rx, below which $E_{0}(l) \leq E_{1}(x)$ for all $x \in (0,l)$, otherwise there at least exists an $x' \in (0,l)$ such that $E_{0}(l) > E_{1}(x')$.
\end{definition}
From the definition, the direct transmission is the optimal when $l \le l_{\text{OP}}$ while the two-hop relay transmission becomes the best (i.e., a relay should be deployed) when $l > l_{\text{OP}}$. As a consequence, the problem becomes how to calculate $l_{\text{OP}}$ and find the optimal position $x_0$ among all $x'$ that achieves $ \min_x E_{1}(x)$, which are quantified by the following theorem, proved in Section~\ref{Section:Solution}.
\begin{theorem} \label{Theorem:OpenDistance}
For each combination of the receive power $P_{R}$ and  target SNR, there exists an open distance $l_{\text{OP}}$, given by
\begin{equation}\label{Eq:OpenDistance}
 l_{\text{OP}} = \max\Bigg\{2\sqrt[\gamma]{\frac{P_{R}\lambda(1-\lambda)}{\psi(\lambda+\gamma)(\lambda+\gamma-1)}},\sqrt[\gamma]{\frac{P_{R}(2-2^{\lambda})}{\psi(2^{\lambda}-2^{1-\gamma})}}\Bigg\}.
\end{equation}
That is, $E_{1}(x)$ achieves its minimum at $x = 0^+$ or $x = l^-$ if $l \le l_{\text{OP}}$. Furthermore, $x_0 = \frac{l}{2}$ is the optimal deploying position if $l > l_{\text{OP}}$, i.e., $ E_{1}\left( \frac{l}{2} \right) =  \min_x E_{1}(x)$.
\end{theorem}

Observe (\ref{Eq:OpenDistance}), $l_{\text{OP}}$ is affected by three factors: devices' restrictions on the receive power $P_{R}$, reliable transmission requirements on the target SNR reflected by $\psi$, and underwater acoustic channels reflected by $\lambda$ and $\gamma$. Figs.~\ref{Fig:TheoOpenDistance}(a) and \ref{Fig:TheoOpenDistance}(b) further intuitively show how these factors affect $l_{\text{OP}}$.

\begin{remark}\label{Remark:OPvsPrSNR}
From (\ref{Eq:OpenDistance}) and Fig.~\ref{Fig:TheoOpenDistance}, two important properties regarding the open distance can be obtained, both of which are totally in accordance with our intuition and will be further verified by simulation results in Section~\ref{Section:Numeric Results}. First, $l_{\text{OP}}$ increases as $P_{R}$ grows for a given target SNR. This is because a larger $P_{R}$ implies that more energy cost is needed to deploy a relay, which can be canceled by the exponentially increased transmit power and decreased bandwidth (see Figs.~\ref{Fig:Bandwidth} and \ref{Fig:PT}) only when the transmission range becomes longer. Second, $l_{\text{OP}}$ decreases as the target SNR increases for a given $P_{R}$, following the fact that transmit power should be increased to satisfy a larger target SNR, but it can be offset by shortening the distance.
\end{remark}

Further checking Fig.~\ref{Fig:TheoOpenDistance} and (\ref{Eq:OpenDistance}), we can find that $l_{\text{OP}}$  is linearly dependent on $P_{R}$ and $\rm{SNR_0}$ on the logarithm scale. Specifically, through a power operation and a logarithm operation on (\ref{Eq:OpenDistance}), it can be equivalently recast to
\begin{equation}\label{Eq:Fitting1}
\begin{aligned}
\gamma \log_{10} l_{\text{OP}} = \log_{10} P_{R} + 0.1\rm{SNR_0} + c
\end{aligned}
\end{equation}
where
\begin{align}
c = \log_{10}\frac{2^{\gamma}\eta\lambda(1-\lambda)}{(\lambda+\gamma)(\lambda+\gamma-1)10^{-4.9528}}
\end{align}
if $l_{\text{OP}} = \sqrt[\gamma]{\frac{P_{R}\lambda(1-\lambda)}{\psi(\lambda+\gamma)(\lambda+\gamma-1)}}$ or
\begin{align}
c = \log_{10}\frac{\eta(2-2^{\lambda})}{(2^{\lambda}-2^{1-\gamma})10^{-4.9528}}
\end{align}
if $l_{\text{OP}} = \sqrt[\gamma]{\frac{P_{R}(2-2^{\lambda})}{\psi(2^{\lambda}-2^{1-\gamma})}}$. From (\ref{Eq:Fitting1}), $\log_{10} l_{\text{OP}}$ is a linear function with respect to both $\log_{10} P_{R}$ and $\rm{SNR_0}$. This implies that $l_{\text{OP}}(P_{R}, \rm{SNR_0})$ can be approximated through polynomial fittings by collecting sufficient  points ($\log_{10} P_{R}$, $\rm{SNR_0}$, $\log_{10} l_{\text{OP}}$), which will be verified in Section~\ref{Subsection:FitOfOD}.

\subsection{Summary: Direct or Relayed?}\label{Subsection:Summary}
Jointly considering Theorem~\ref{Theorem:OpenDistance} and Eq.~(\ref{Eq:limitValue}), we have
\begin{equation}\label{Eq:Solution}
\min_x E_{1}(x) = \left\{
\begin{aligned}
& E_{1}(0^{+}) = E_{1}(l^{-}) = E_{0}(l), \text{when} \ l \leq l_{\text{OP}}\\
& E_{1}(l/2), \text{when} \ l > l_{\text{OP}}
\end{aligned}
\right.
\end{equation}
where $l_{\text{OP}}$ is given by (\ref{Eq:OpenDistance}). Eq.~(\ref{Eq:Solution}) quantitatively answers the two questions proposed in Section~\ref{Subsection:ProblemFormulation}. More clearly,
\begin{itemize}
  \item A relay should not be introduced when $l \leq l_{\text{OP}}$, i.e., direct transmission is the best.
  \item A relay should be introduced when $l > l_{\text{OP}}$ (i.e., two-hop relay transmission is the optimal) and its optimal placing position is the midpoint between the Rx and Tx.
  \item After a relay is deployed at the midpoint, another question that arises is whether more relays are needed to further cut down the energy expenditure, which can be easily decided utilizing Theorem~\ref{Theorem:OpenDistance}. Specifically, two another relays need to be deployed at the quarter points if  $l/2 > l_{\text{OP}}$ and four more additional relays are further needed at the eighth points if $l/4 > l_{\text{OP}}$. This process is terminated until the transmission distance of each hop is less than $l_{\text{OP}}$.
\end{itemize}

\section{Solution Proof and Analysis} \label{Section:Solution}
In this section, we strictly elaborate how Theorem~\ref{Theorem:OpenDistance} is obtained, where the differential analysis is adopted. We first show the bounded and continuous features of $E_{1}(x)$, which is the pre-condition of using the differential analysis. We then derive the sign of $\frac{d E_{1}^{3}(x)}{d x^{3}}$, $\frac{d E_{1}^{2}(x)}{d x^{2}}$, and $\frac{d E_{1}(x)}{d x}$ to obtain the rough figure of $E_{1}(x)$, based on which the locally extreme and globally optimal points are found.

\subsection{Preliminary}\label{Subsection:Methodology}
In the paper, we adopt the differential analysis to find the global minimum of $E_{1}(x)$. Since not all functions are suitable for differential analysis, we first present the following lemma, which directly follows the fact that $E_{1}(x)$ is a linear combination of several power functions, to support the utilization of this theory.

\begin{lemma} \label{Lemma:continuity}
In the definition field $x\in (0,l)$, $E_{1}(x)$ is a bounded and continuous function with respect to $x$  and its $n$-order ($n \in \mathbb{N^{+}}$) derivatives make sense.
\end{lemma}

Lemma~\ref{Lemma:continuity} indicates that we can decide the global minimum of $E_{1}(x)$ by first finding out all its extreme points based on the differential analysis and then comparing their objective values with those achieved by the boundary points (i.e., $x=0^{+}$ and $x=l^{-}$). The following lemma further shows the sign of the second derivative of $E_{1}(x)$, which facilitates us to judge whether a stationary point (where $\frac{d E_{1}(x)}{dx}=0$) is an extreme point or not and will be used in the next two subsections.
\begin{lemma} \label{Lemma:2ndDerivative}
The second derivative of $E_{1}(x)$ with respect to $x$ for $x\in (0,l)$ satisfies the following inequality
    \begin{equation}\label{Eq:Condition}
    \begin{aligned}
      \frac{d E_{1}^{2}(x)}{d x^{2}} &\leq  \frac{2L}{\alpha \omega}(l/2)^{\lambda-2}\big[\psi(\lambda+\gamma)(\lambda+\gamma-1)(l/2)^{\gamma} \\
      & \ \ \ - P_{R}\lambda(1-\lambda)\big], \forall x\in (0,l).
    \end{aligned}
    \end{equation}
\end{lemma}

\begin{IEEEproof}
From (\ref{Eq:RelayNetworksEnergy}), the first derivative of $E_{1}$ with respect to $x$ is given by
    \begin{equation}\label{Eq:FirstD}
    \begin{aligned}
    \frac{dE_{1}(x)}{dx} &= \ \frac{L}{\alpha \omega}\Big\{\psi(\lambda+\gamma)\left[x^{\lambda+\gamma-1} - (l-x)^{\lambda+\gamma-1}\right] +\\
    & \ \ \ \ \ \  \ \ \ \ P_{R}\lambda\left[x^{\lambda-1} - (l-x)^{\lambda-1}\right]\Big\}, x\in (0,l).
    \end{aligned}
    \end{equation}
Initial purpose is to find some $x^*$ that satisfies $\frac{dE_{1}(x^*)}{dx} = 0$. However, it is hard to judge whether (\ref{Eq:FirstD}) is positive or negative. Nevertheless, we can still ensure the values of some key points
   \begin{equation}\label{Eq:SomeFirstD}
   \begin{aligned}
   \left. \frac{d E_{1}(x)}{d x}\right|_{x=0^{+}} &\rightarrow \infty\\
   \left. \frac{d E_{1}(x)}{d x}\right|_{x=l^{-}} &\rightarrow -\infty\\
   \left. \frac{d E_{1}(x)}{d x}\right|_{x=l/2} &= 0.
   \end{aligned}
   \end{equation}

Furthermore, we calculate the second derivative of $E_{1}$ with respect to $x$ in $x\in (0,l)$
   \begin{equation}\label{Eq:SecondD}
   \begin{aligned}
   &\frac{d E_{1}^{2}(x)}{d x^{2}} = \frac{L}{\alpha \omega}\Big\{-P_{R}\lambda(1-\lambda)\left[x^{\lambda-2} + (l-x)^{\lambda-2}\right]\\
   &+\psi(\lambda+\gamma)(\lambda+\gamma-1)\left[x^{\lambda+\gamma-2}+(l-x)^{\lambda+\gamma-2}\right]\Big\}.\\
   \end{aligned}
   \end{equation}
Since $\lambda+\gamma-2\in(0,1)$ and $\lambda-2 \in (-2,-1)$ from (\ref{Eq:parameterScale}),  $x^{\lambda+\gamma-2}$  and $x^{\lambda-2}$ are thus a concave function and a convex function, respectively, when $x\in(0,l)$.  As a consequence, we can apply the Jensen's inequality to obtain the following inequalities
   \begin{equation}\label{Eq:Jensen}
   \begin{aligned}
   x^{\lambda+\gamma-2} + (l-x)^{\lambda+\gamma-2} &\leq 2 \cdot (l/2)^{\lambda+\gamma-2}\\
   x^{\lambda-2} + (l-x)^{\lambda-2} &\geq 2 \cdot (l/2)^{\lambda-2}.
   \end{aligned}
   \end{equation}
Substituting (\ref{Eq:Jensen}) into (\ref{Eq:SecondD}), we can shrink the range of the second derivative to obtain (\ref{Eq:Condition}).
\end{IEEEproof}

For notational simplicity, we set $D = \frac{2L}{\alpha \omega}(l/2)^{\lambda-2}\big[\psi(\lambda+\gamma)(\lambda+\gamma-1)(l/2)^{\gamma} - P_{R}\lambda(1-\lambda)\big]$. From Lemma~\ref{Lemma:2ndDerivative}, the sign of $\frac{d E_{1}^{2}(x)}{d x^{2}}$ depends on two cases of the positivity or negativity of $D$, which affects the optimal solution of $E_{1}(x)$. The next two subsections discuss how to identify the extreme points of $E_{1}(x)$ in each of these two cases, respectively.

\begin{figure}[t]
\centering \leavevmode \epsfxsize=3.5 in  \epsfbox{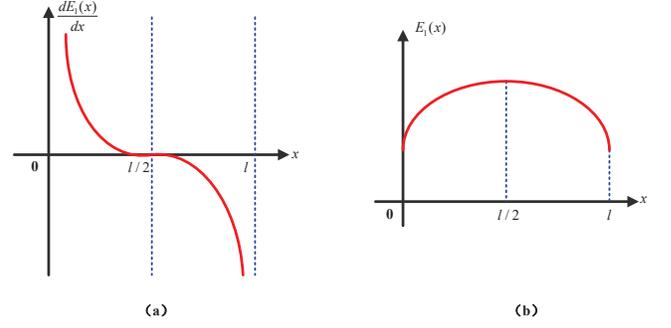}
\centering \caption{Illustration of the variation trend of $\frac{d E_{1}(x)}{d x}$ and $E_{1}$ with $x$. (a) $\frac{d E_{1}(x)}{d x}$ vs. $x$. (b) $E_{1}(x)$ vs. $x$.
 } \label{Fig:case1}
\end{figure}

\subsection{Case 1: $D \le 0$}\label{Subsection:OptimalValue}
In this case, $\frac{d E_{1}^{2}(x)}{d x^{2}} \leq 0$ holds for all $x \in (0,l)$ from (\ref{Eq:Condition}). Based on this fact, we can prove that a stationary point of $E_{1}(x)$ is the globally maximum point and the solution of (\ref{Eq:Problem_Formulation}) is reached at the two boundary points $0^{+}$ or $l^{-}$. The following theorem quantifies these results.

\begin{theorem} \label{Theorem:Case1}
$E_{1}(x)$ gets its global minimum at $x=0^{+}$ or at $x=l^{-}$ when $l \leq 2\sqrt[\gamma]{\frac{P_{R}\lambda(1-\lambda)}{\psi(\lambda+\gamma)(\lambda+\gamma-1)}}$.
\end{theorem}

\begin{IEEEproof}
Rearranging $l \leq 2\sqrt[\gamma]{\frac{P_{R}\lambda(1-\lambda)}{\psi(\lambda+\gamma)(\lambda+\gamma-1)}}$ yields
\begin{equation}\label{Eq:Case1}
\begin{aligned}
&\psi(\lambda+\gamma)(\lambda+\gamma-1)(l/2)^{\gamma} - P_{R}\lambda(1-\lambda) \leq 0.
\end{aligned}
\end{equation}
From Lemma~\ref{Lemma:2ndDerivative} (see (\ref{Eq:Condition})), $\frac{d E_{1}^{2}(x)}{d x^{2}} \leq 0$ holds for all $x \in (0,l)$ under this condition. Then taking (\ref{Eq:SomeFirstD}) into account, we can roughly draw the variation trend of $\frac{d E_{1}(x)}{d x}$ with $x$, as shown in Fig.~\ref{Fig:case1}(a). Further considering the positivity and symmetry properties of $E_{1}(x)$, we sketch how $E_{1}(x)$ varies with $x$ in Fig.~\ref{Fig:case1}(b). Specifically, $E_{1}(x)$ first increases from $x=0^{+}$ to $x=l/2$ and then decreases from $x=l/2$ to $x=l^{-}$, and thus $x=l/2$ can be ensured as the only extreme point as well as the globally maximum point for $x\in (0,l)$. Moreover, from (\ref{Eq:limitValue}), we can obtain
  \begin{equation}\label{Eq:OpenD1}
  \begin{aligned}
  &\{E_{1}(x)\}_{\min} = E_{1}(x=0^{+}) = E_{1}(x=l^{-}) = E_{0}(l)\\
  &\Rightarrow E_{0}(l) \leq E_{1}(x), \forall x\in (0,l).
  \end{aligned}
  \end{equation}
We complete the proof of Theorem~\ref{Theorem:Case1}.
\end{IEEEproof}

\begin{remark}
Theorem~\ref{Theorem:Case1} indicates that, once the transmission range $l \le 2\sqrt[\gamma]{\frac{P_{R}\lambda(1-\lambda)}{\psi(\lambda+\gamma)(\lambda+\gamma-1)}}$, the energy consumption of direct transmission is always less than that of two-hop relay transmission, i.e., a relay should not be deployed in this case. In addition, the maximum overall energy cost is reached when the relay is deployed at the midpoint of the link.
\end{remark}

\subsection{Case 2: $D > 0$}

In this case, $\frac{d E_{1}^{2}(x)}{d x^{2}} \leq 0$ possibly does not hold for some $x \in (0,l)$ (see (\ref{Eq:Condition})), and thus a stationary point of $E_{1}(x)$ may not be an extreme point. In spite of this, we can infer how $E_{1}(x)$ varies with $x$ successively from the features of $\frac{d E_{1}^{3}(x)}{d x^{3}}$,$\frac{d E_{1}^{2}(x)}{d x^{2}}$, and $\frac{d E_{1}(x)}{d x}$. The following lemma, which will be used in Theorem \ref{Theorem:Case2} and Theorem \ref{Theorem:Case3}, quantifies all the possible optimal solutions of (\ref{Eq:Problem_Formulation}) in this case.

\begin{lemma} \label{Lemma:3D}
$E_{1}(x)$ gets its global minimum at $x=l/2$, $x=0^{+}$, or $x=l^{-}$ when $l > 2\sqrt[\gamma]{\frac{P_{R}\lambda(1-\lambda)}{\psi(\lambda+\gamma)(\lambda+\gamma-1)}}$.
\end{lemma}

\begin{figure}[t]
\centering \leavevmode \epsfxsize=3.5 in  \epsfbox{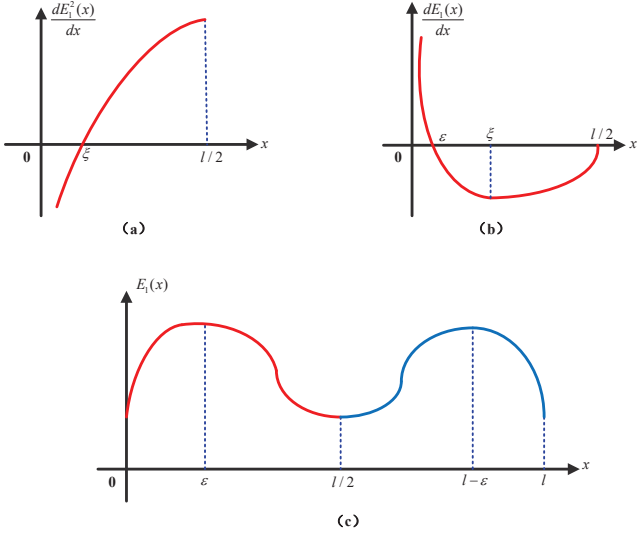}
\centering \caption{Illustration of the variation trend of $\frac{d E_{1}^{2}(x)}{d x^{2}}$, $\frac{d E_{1}(x)}{d x}$, and $E_{1}$ with $x$. (a) $\frac{d E_{1}^{2}(x)}{d x^{2}}$ vs. $x$. (b) $\frac{d E_{1}(x)}{d x}$ vs. $x$. (c) $E_{1}$ vs. $x$.} \label{Fig:case2}
\end{figure}

\begin{IEEEproof}
Similarly, rearranging $l > 2\sqrt[\gamma]{\frac{P_{R}\lambda(1-\lambda)}{\psi(\lambda+\gamma)(\lambda+\gamma-1)}}$ yields
  \begin{equation}\label{Eq:Case2}
  \psi(\lambda+\gamma)(\lambda+\gamma-1)(l/2)^{\gamma} - P_{R}\lambda(1-\lambda) > 0.
  \end{equation}
From Lemma~\ref{Lemma:2ndDerivative}$, \frac{d E_{1}^{2}(x)}{d x^{2}} \leq 0$ possibly does not hold for some $x \in (0,l)$ under this condition (see (\ref{Eq:Condition})). Still, the values of some key points regarding $\frac{d E_{1}^{2}(x)}{d x^{2}}$ can be identified from (\ref{Eq:SecondD})
  \begin{equation}\label{Eq:SomeSecondD}
  \begin{aligned}
  &\left. \frac{d E_{1}^{2}(x)}{d x^{2}}\right|_{x=0^{+}} \rightarrow -\infty\\
  &\left. \frac{d E_{1}^{2}(x)}{d x^{2}}\right|_{x=l^{-}} \rightarrow -\infty\\
  \left. \frac{d E_{1}^{2}(x)}{d x^{2}}\right|_{x=l/2} &= (l/2)^{\lambda-2}\frac{2L}{\alpha \omega}\big[\psi(\lambda+\gamma)(\lambda+\gamma-1)(l/2)^{\gamma}\\
  &\ \ \ \ \ - P_{R}\lambda(1-\lambda)\big] > 0.
  \end{aligned}
  \end{equation}

Furthermore, we compute the third derivative of $E_{1}$ with respect to $x$ as
 \begin{equation}\label{Eq:ThirdD}
 \begin{aligned}
  \frac{d E_{1}^{3}(x)}{d x^{3}} &= \frac{L}{\alpha \omega}\Big\{P_{R}\lambda(1-\lambda)(2-\lambda)\left[x^{\lambda-3} - (l-x)^{\lambda-3}\right]\\
  & \ \ \ +\psi(\lambda+\gamma)(\lambda+\gamma-1)(\lambda+\gamma-2) \cdot\\
  & \ \ \ \ \big[x^{\lambda+\gamma-3}-(l-x)^{\lambda+\gamma-3}\big]\Big\}, x\in (0,l)
 \end{aligned}
 \end{equation}
where $\lambda-3 \in (-3,-2)$ and ${\lambda+\gamma-3} \in (-1,0)$ from (\ref{Eq:parameterScale}). Notice that $E_{1}(x) = E_{1}(l-x)$, so that we just need to consider the semi-open interval $(0,l/2]$, and the other half can be derived from the symmetry. Moreover, as $\lambda-3$ and $\lambda+\gamma-3$ are both negative, $x^{\lambda-3}$ and $x^{\lambda+\gamma-3}$ thus decrease from $x=0^{+}$ to $x=l^-$, which implies the following relations
 \begin{equation}\label{Eq:Third0}
 \begin{aligned}
  x^{\lambda-3} - (l-x)^{\lambda-3} &\geq 0, x\in (0,l/2]\\
  x^{\lambda+\gamma-3}-(l-x)^{\lambda+\gamma-3} &\geq 0, x\in (0,l/2].
 \end{aligned}
 \end{equation}
Substituting (\ref{Eq:Third0}) into (\ref{Eq:ThirdD}) yields
\begin{equation}
\frac{d E_{1}^{3}(x)}{d x^{3}} \geq 0, \forall x\in (0,l/2].
\end{equation}
Combined with (\ref{Eq:SomeSecondD}), the sign of $\frac{d E_{1}^{2}(x)}{d x^{2}}$ for $x \in (0,l/2)$ can be determined as
 \begin{equation}\label{Eq:Second0}
  \frac{d E_{1}^{2}(x)}{d x^{2}} \left\{
  \begin{aligned}
   &< 0, x\in (0,\xi)\\
   &= 0, x = \xi \\
   &> 0, x\in (\xi,l/2]
  \end{aligned}
  \right.
 \end{equation}
where $\xi$ is a positive constant less than $l/2$. Fig. \ref{Fig:case2}(a) sketches the rough curve of $\frac{d E_{1}^{2}(x)}{d x^{2}} $ with $x$.

Based on (\ref{Eq:SomeFirstD}), we can further confirm the sign of $\frac{d E_{1}(x)}{d x}$ as follows
\begin{equation}\label{Eq:First0}
  \frac{d E_{1}(x)}{d x} \left\{
  \begin{aligned}
   &> 0, x\in (0,\epsilon)\\
   &= 0, x = \epsilon \\
   &< 0, x\in (\epsilon,l/2]
  \end{aligned}
 \right.
\end{equation}
where $\epsilon$ is a positive constant less than $\xi$, and we plot $\frac{d E_{1}(x)}{d x}$ in Fig.~\ref{Fig:case2}(b). At this point, how $E_{1}(x)$ varies with $x$ in $(0,l/2]$ becomes clear, which is displayed in Fig.~\ref{Fig:case2}(c). More specifically, $E_{1}(x)$ first keeps going up from $x=0^{+}$ to $x=\epsilon$ and then going down from $x=\epsilon$ to $x=l/2$. According to the symmetric feature of $E_{1}(x)$, we can easily obtain the rough figure of $E_{1}(x)$ when $x \in [l/2,l)$ from that of $x \in (0,l/2]$. From the figure, $x=l/2$ is the only locally minimum point while $x=\epsilon$ and $x=l-\epsilon$ are two locally maximum points of $E_{1}(x)$. Hence, $E_{1}(x)$ gets the minimum value at $x = l/2$, $x= 0^{+}$, or $x = l^{-}$.
\end{IEEEproof}

Since $E_1(0^{+}) = E_1(l^{-})$, Lemma~\ref{Lemma:3D} indicates that there are two cases regarding the optimal solution of (\ref{Eq:Problem_Formulation}) when $l > 2\sqrt[\gamma]{\frac{P_{R}\lambda(1-\lambda)}{\psi(\lambda+\gamma)(\lambda+\gamma-1)}}$, i.e., it is achieved at $x = 0^{+}$ and $x = l^{-}$ or at $x = l/2$. Theorems~\ref{Theorem:Case2} and \ref{Theorem:Case3} further specify these two cases, respectively.

\begin{theorem} \label{Theorem:Case2}
$E_{1}(x)$ gets its global minimum at $x=0^{+}$ or $x=l^{-}$ when $l \leq \sqrt[\gamma]{\frac{P_{R}(2-2^{\lambda})}{\psi(2^{\lambda}-2^{1-\gamma})}}$.
\end{theorem}

\begin{IEEEproof}
We know that (\ref{Eq:Problem_Formulation}) gets its optimal at $x=0^{+}$ or at $x=l^{-}$ if $l \le \sqrt[\gamma]{\frac{P_{R}(2-2^{\lambda})}{\psi(2^{\lambda}-2^{1-\gamma})}} \leq 2\sqrt[\gamma]{\frac{P_{R}\lambda(1-\lambda)}{\psi(\lambda+\gamma)(\lambda+\gamma-1)}}$ from Theorem~\ref{Theorem:Case1}.
However, when $2\sqrt[\gamma]{\frac{P_{R}\lambda(1-\lambda)}{\psi(\lambda+\gamma)(\lambda+\gamma-1)}} < l \leq \sqrt[\gamma]{\frac{P_{R}(2-2^{\lambda})}{\psi(2^{\lambda}-2^{1-\gamma})}}$, $\frac{d E_{1}^{2}(x)}{d x^{2}} \leq 0$ cannot be ensured and thus Theorem~\ref{Theorem:Case1} cannot be applied. In this case, we first rearrange the condition $l \leq \sqrt[\gamma]{\frac{P_{R}(2-2^{\lambda})}{\psi(2^{\lambda}-2^{1-\gamma})}}$ as follows
\begin{equation}\label{Eq:Case21}
\psi l^{\lambda}(2^{\gamma}-2^{1-\lambda}) - P_{R}(2-2^{\gamma}) \leq 0.
\end{equation}
As $l>2\sqrt[\gamma]{\frac{P_{R}\lambda(1-\lambda)}{\psi(\lambda+\gamma)(\lambda+\gamma-1)}}$, $E_{1}(x)$ gets its global minimum at $x=l/2$, $x=0^{+}$, or $x=l^{-}$ from Lemma~\ref{Lemma:3D}. Furthermore,  from (\ref{Eq:Case21}), we have the following inequality
 \begin{equation}\label{Eq:Conclusion2}
 \begin{aligned}
  &E_{1}(0^{+}) - E_{1}(l/2) = E_{1}(l^{-}) - E_{1}(l/2) \\
  &= \frac{L}{\alpha \omega}\big\{\psi l^{\lambda+\gamma} + P_{R}l^{\lambda} - 2\cdot\psi(l/2)^{\lambda+\gamma} - 2\cdot P_{R}(l/2)^{\lambda} \big\} \\
  &= \frac{L}{\alpha \omega}\cdot(l/2)^{\gamma}\big\{\psi l^{\lambda}(2^{\gamma}-2^{1-\lambda}) - P_{R}(2-2^{\gamma}) \big\} \leq 0
 \end{aligned}
 \end{equation}
Eq.~(\ref{Eq:Conclusion2}) implies that $E_{1}(x)$ achieves its minimum also at $x=0^{+}$ or $x=l^{-}$ in the case of $2\sqrt[\gamma]{\frac{P_{R}\lambda(1-\lambda)}{\psi(\lambda+\gamma)(\lambda+\gamma-1)}} < l \leq \sqrt[\gamma]{\frac{P_{R}(2-2^{\lambda})}{\psi(2^{\lambda}-2^{1-\gamma})}}$.
\end{IEEEproof}

\begin{remark}
Theorem \ref{Theorem:Case2} tells that, once the transmission range $l \le \sqrt[\gamma]{\frac{P_{R}(2-2^{\lambda})}{\psi(2^{\lambda}-2^{1-\gamma})}}$, the energy consumption of direct transmission is also always less than that of two-hop relay transmission, i.e., it should not introduce a relay. It is worthwhile to note that, unlike the case in Theorem~\ref{Theorem:Case1}, the overall energy consumption when deploying the relay at the midpoint of the link is neither the highest nor the lowest in this case (see Fig.~\ref{Fig:case2}(c)).
\end{remark}

\begin{theorem} \label{Theorem:Case3}
$E_{1}(x)$ gets its global minimum at $x=l/2$ when $l > \max\Bigg\{2\sqrt[\gamma]{\frac{P_{R}\lambda(1-\lambda)}{\psi(\lambda+\gamma)(\lambda+\gamma-1)}},\sqrt[\gamma]{\frac{P_{R}(2-2^{\lambda})}{\psi(2^{\lambda}-2^{1-\gamma})}}\Bigg\}$.
\end{theorem}

\begin{IEEEproof}
Recall that, when $l > 2\sqrt[\gamma]{\frac{P_{R}\lambda(1-\lambda)}{\psi(\lambda+\gamma)(\lambda+\gamma-1)}}$, $E_{1}(x)$ gets its global minimum at $x=l/2$, $x=0^{+}$, or $x=l^{-}$ from Lemma~\ref{Lemma:3D}. Rearranging $l > \sqrt[\gamma]{\frac{P_{R}(2-2^{\lambda})}{\psi(2^{\lambda}-2^{1-\gamma})}}$, we obtain
  \begin{equation}\label{Eq:Case22}
  \psi l^{\lambda}(2^{\gamma}-2^{1-\lambda}) - P_{R}(2-2^{\gamma}) > 0.
  \end{equation}
  Leveraging (\ref{Eq:Case22}), the following inequality can be derived.
 \begin{equation}\label{Eq:Conclusion2_Derived}
 \begin{aligned}
  &E_{1}(0^{+}) - E_{1}(l/2) = E_{1}(l^{-}) - E_{1}(l/2) \\
  &= \frac{L}{\alpha \omega}\big\{\psi l^{\lambda+\gamma} + P_{R}l^{\lambda} - 2\cdot\psi(l/2)^{\lambda+\gamma} - 2\cdot P_{R}(l/2)^{\lambda} \big\} \\
  &= \frac{L}{\alpha \omega}\cdot(l/2)^{\gamma}\big\{\psi l^{\lambda}(2^{\gamma}-2^{1-\lambda}) - P_{R}(2-2^{\gamma}) \big\} > 0.
 \end{aligned}
 \end{equation}
Considering these two cases, we can obtain that $E_{1}(x)$ achieves its minimum at $x=l/2$.
\end{IEEEproof}
\begin{remark}
Theorem \ref{Theorem:Case3} indicates that, it is necessary to deploy a relay once $l > \max\Bigg\{2\sqrt[\gamma]{\frac{P_{R}\lambda(1-\lambda)}{\psi(\lambda+\gamma)(\lambda+\gamma-1)}},\sqrt[\gamma]{\frac{P_{R}(2-2^{\lambda})}{\psi(2^{\lambda}-2^{1-\gamma})}}\Bigg\}$ and the relay should be deployed at the middle point between the Rx and the Tx to achieve the smallest energy consumption.
\end{remark}

To summarize, Theorem~\ref{Theorem:OpenDistance} is obtained by jointly considering Theorems~\ref{Theorem:Case1}, \ref{Theorem:Case2}, and \ref{Theorem:Case3}.

\begin{figure}[t]
\centering \leavevmode \epsfxsize=3.5 in  \epsfbox{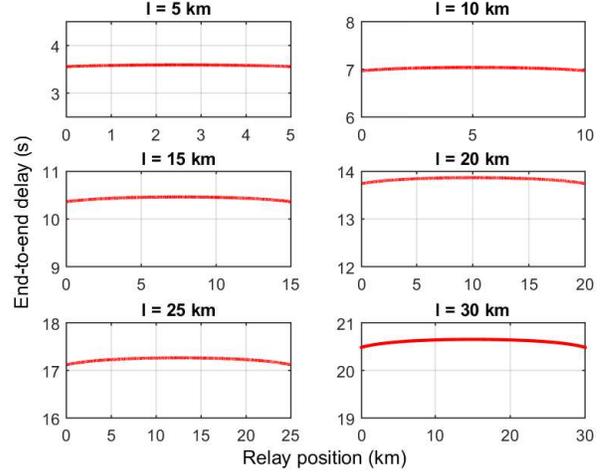}
\centering \caption{Examples of the end-to-end delay with respect to relay positions under different transmission distance settings, where the target SNR and the receive power are set to be 15 dB and 1.0 W, respectively.} \label{Fig:ARelayDelay}
\end{figure}
\begin{figure}[t]
\centering \leavevmode \epsfxsize=3.5 in  \epsfbox{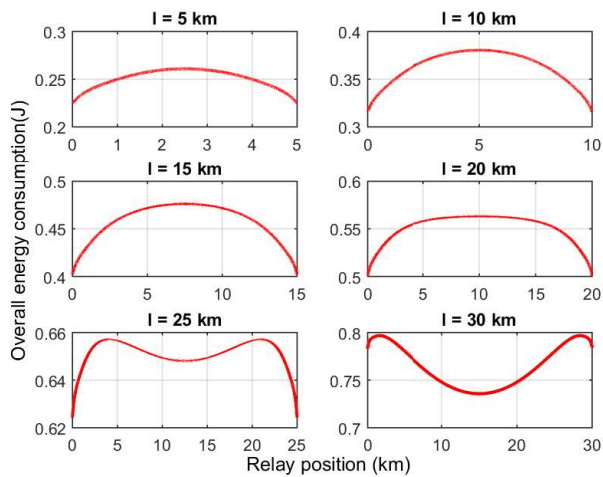}
\centering \caption{Examples of the overall energy consumption with respect to relay positions different transmission distance settings, where the target SNR and the receive power are set to be 15 dB and 1.0 W, respectively.} \label{Fig:ARelayEnergy}
\end{figure}

\begin{table*}[tp]
  \centering
  \caption{Performance comparison between direct transmission and equidistant two-hop relay transmission.}
  \label{tab:performance_comparison}
    ~~~~~~\begin{tabular}{|c|c|c|c|c|c|c|c|}
    \hline
    \multirow{2}{*}{$\rm{SNR}_0$/dB}&\multirow{2}{*}{$l$/km}&
    \multicolumn{6}{c|}{$P_{R}=0.5$ W}\cr\cline{3-8}
    &&$E_{0}(l)$/J&$E_{1}(l/2)$/J&Reduction ratio: $\frac{E_{0}(l)-E_{1}(l/2)}{E_{0}(l)}$&$D_{0}(l)$/s&$D_{1}(l/2)$/s& Reduction ratio: $\frac{D_{0}(l)-D_{1}(l/2)}{D_{0}(l)}$\cr
    \hline
    10&10&0.1381&0.1893&-27.05\%&6.9338 &7.0423 &-1.56\%\cr\hline
    10&20&0.2157&0.2762&-28.05\%&13.7048&13.8675&-1.19\%\cr\hline
    10&30&0.3198&0.3521&-10.10\%&20.4450&20.6500&-1.00\%\cr\hline
    10&40&0.4808&0.4351&9.50\%  &27.1769&27.4095&-0.84\%\cr\hline
    10&50&0.7274&0.5238&27.99\% &33.9107&34.1540&-0.72\%\cr\hline
    15&10&0.1480&0.1926&-32.57\%&6.9338 &7.0423 &-1.56\%\cr\hline
    15&20&0.2807&0.2960&-5.45\% &13.7048&13.8675&-1.19\%\cr\hline
    15&30&0.5275&0.4107&22.14\% &20.4450&20.6500&-1.00\%\cr\hline
    15&40&0.9689&0.5614&42.06\% &27.1769&27.4095&-0.84\%\cr\hline
    15&50&1.6759&0.7691&54.11\% &33.9107&34.1540&-0.72\%\cr\hline
    20&10&0.1793&0.2030&-13.21\%&6.9338 &7.0423 &-1.56\%\cr\hline
    20&20&0.4861&0.3585&26.25\% &13.7048&13.8675&-1.19\%\cr\hline
    20&30&1.1869&0.5960&49.78\% &20.4450&20.6500&-1.00\%\cr\hline
    20&40&2.5124&0.9722&61.30\% &27.1769&27.4095&-0.84\%\cr\hline
    20&50&4.6755&1.5448&66.96\% &33.9107&34.1540 &-0.72\%\cr\hline
    25&10&0.2781&0.2358&15.21\%&6.9338 &7.0423 &-1.56\%\cr\hline
    25&20&1.1356&0.5562&51.02\% &13.7048&13.8675&-1.19\%\cr\hline
    25&30&3.2721&1.1820&63.88\% &20.4450&20.6500&-1.00\%\cr\hline
    25&40&7.3933&2.2713&69.27\% &27.1769&27.4095&-0.84\%\cr\hline
    25&50&14.1609&3.9979&71.77\% &33.9107&34.1540 &-0.72\%\cr\hline
    \end{tabular}
\end{table*}

\section{Simulation Results and Analysis} \label{Section:Numeric Results}
In this section, we present extensive simulation results to exhibit the performance of relay-aided underwater acoustic networks as well as to verify our derived theoretical results in Theorems~\ref{Theorem:OpenDistance}--~\ref{Theorem:Case3}. We adopt Fig.~\ref{Fig:Scenario} as the simulation scenario, where the packet size $L=256$ Bytes and BPSK is chosen as the modulation method (and thus the bandwidth efficiency $\alpha=1$). Typical parameter settings provided in \cite{UnderwaterAcousticCommunication_CapacityVSDistance_ACMSIGMOBILE2007,UnderwaterAcousticCommmunication_ARelayScenarios_Oceans2007,
UnderwaterAcousticNetworks_EnergyEfficientRouting_ieeeJSAC2008} are used throughout the simulation results. Specifically, the overall energy efficiency of the electronic circuitry $\eta=0.25$, moderate shipping activity ($s=0.5$) and no wind ($w=0$) are set for the noise p.s.d. (\ref{Eq:NoiseComponent}), and practical spreading ($k=1.5$) is used for the path loss model (\ref{Eq:PathLoss}).

\subsection{End-to-End Delay and Overall Energy Consumption}\label{Subsection:NetworkPerformance}
In Fig.~\ref{Fig:ARelayDelay} and Fig.~\ref{Fig:ARelayEnergy}, we show how the relay position affects the network performance in terms of the end-to-end delay and the overall energy consumption under different transmission distances. Specifically, Fig.~\ref{Fig:ARelayDelay} exhibits that the end-to-end delay in all cases is almost invariant with the relay position, this is because it is dominated by the propagation delay due to the low sound speed (about 1500 m/s underwater).
From Fig.~\ref{Fig:ARelayEnergy}, all the overall energy consumption curves in cases of $l=5$--20 km have their maximal values at the corresponding midpoints and reach their minimum at $x=0^{+}$ or $x=l^{-}$, which validates the correctness of Theorem~\ref{Theorem:Case1}. On the contrary, there are two local maximum points and one local minimum points when $l=25$ km and $l=30$ km, and the minimum of the overall energy consumption is achieved at $x=15$ km (when $l=30$ km) or at $x=0^{+}$ and $x=l^{-}$ (when $l=25$ km), which verify the correctness of Theorem~\ref{Theorem:Case3} and Theorem~\ref{Theorem:Case2}, respectively.

Furthermore, we quantitatively compare the network performance of direct transmission with that of equidistant two-hop relay transmission in Table~\ref{tab:performance_comparison}. It is observed from the table that, properly (i.e., in the case when the transmission distance is larger than the open distance) deploying a relay at the middle point can dramatically reduce the network energy consumption but almost without increasing the end-to-end delay. For example, the energy expenditure is saved by up to 71.77\% but with only a 0.72\% increase in the delay when $l = 50$ km and $\rm{SNR_0}$= 25 dB. Moreover, these data shows that the performance improvement of two-hop relay transmission will become higher with the growth of transmission distance and target SNRs.

\subsection{Open Distance and Error Analysis}\label{Subsection:ExistenceOfOD}
\begin{figure}[t]
\centering \leavevmode \epsfxsize=3.5 in  \epsfbox{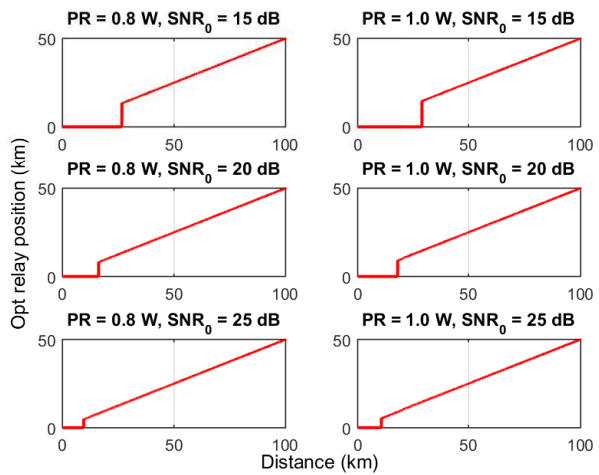}
\centering \caption{Optimal relay position under different combinations of the target SNR and receive power to minimize the overall energy consumption of two-hop relay underwater acoustic links.} \label{Fig:optimalRelayPosition}
\end{figure}
In order to verify the existence and properties of the open distance, we also plot the realistic results\footnote{Note that realistic results/values in this section are obtained by directly substituting (\ref{Eq:PathLoss})--(\ref{Pt}) into (\ref{Eq:Problem_Formulation}) without any simplification or approximation (i.e., not using the fitting expressions in (\ref{Eq:PT}) and (\ref{Eq:uPaToWatt})) and solving (\ref{Eq:Problem_Formulation}) through exhaustive search over $x \in [0,l]$.} of the optimal relay position (where the overall energy consumption reaches its minimum) by exhaustive search for each combination of the target SNR and receive power $P_R$  in Fig.~\ref{Fig:optimalRelayPosition}. Observe that, for all of the six combinations, the optimal relay position firstly remains at 0 km (i.e, deploying a relay is not needed), and then keeps increasing linearly at a slope of 0.5 (which means that the relay should be located at the midpoint of the transmission link) once the transmission distance exceeds a certain value. These phenomenons prove the existence of the open distance and the correctness of Theorem~\ref{Theorem:OpenDistance}. In addition, as exhibited in Fig.~\ref{Fig:TheoOpenDistance} and discussed in Remark~\ref{Remark:OPvsPrSNR}, Fig.~\ref{Fig:optimalRelayPosition} also shows that the open distance grows with the receive power and decreases with the target SNR.

\begin{figure}[t]
\centering \leavevmode \epsfxsize=3.5 in  \epsfbox{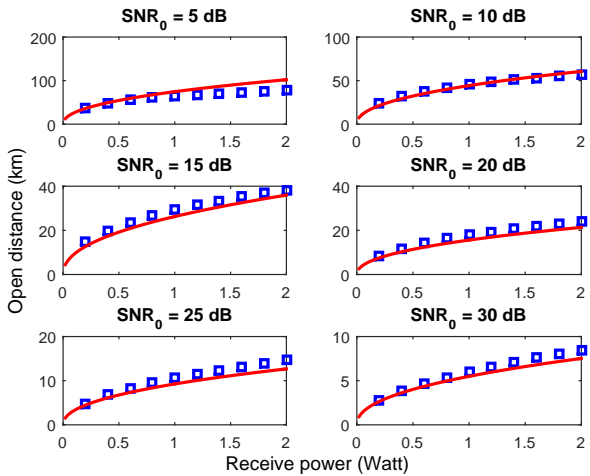}
\centering \caption{Comparison between the theoretical and realistic values for the open distance under different target SNR settings, where boxes represent realistic values and curves denote theoretical results. In the figure, we set $\lambda =0.5392$ and $\gamma = 2.2074$.} \label{Fig:error}
\end{figure}

Furthermore, we extract the turning points of the curves (i.e., realistic open distance) in Fig.~\ref{Fig:optimalRelayPosition} for each combination of the target SNR and receive power $P_R$ and compare them with the theoretical open distance in Fig.~\ref{Fig:error}. The figure exhibits that the theoretical curves overall match well with the realistic values under different target SNR settings, which verifies the validity of (\ref{Eq:OpenDistance}). In the meantime, the obtained results show that their deviations also exist, particularly in some cases with low target SNRs, which are mainly due to the following two factors. First, the modeling error incurred by the numerical approximation for the effective bandwidth (\ref{Eq:PT}) and the required transmit power (\ref{Eq:uPaToWatt}). Second, the roundoff error resulted from a series of power operations in (\ref{Eq:OpenDistance}). In a nutshell, simulation results reveal that (\ref{Eq:OpenDistance}) matches well with realistic results in the target SNR and receive power ranges of our interest.

\begin{table}[t]
\centering
\caption{Evaluation of the polynomial fitting (\ref{Eq:ApproximationExpression}) in terms of the GoF indexes.}\label{Table:FittingResults}
\begin{supertabular}{|c|c|c|c|c|c|}
\hline
m & n & SSE & RMSE & $R^{2}$ & adj-$R^{2}$ \\
\hline
1 & 1 & 3.613 & 0.02803 & 0.998 & 0.998 \\
\hline
2 & 2 & 2.887 & 0.02507 &  0.9984 &  0.9984  \\
\hline
3 & 3 & 0.2969 & 0.008042 & 0.9998 & 0.9998  \\
\hline
4 & 4 & 0.1153 & 0.005014 & 0.9999 & 0.9999 \\
\hline
5 & 5 & 0.03683 & 0.002836 & 1.0000 & 1.0000  \\
\hline
\end{supertabular}
\end{table}

\begin{figure}[t]
\centering \leavevmode \epsfxsize=3.5 in  \epsfbox{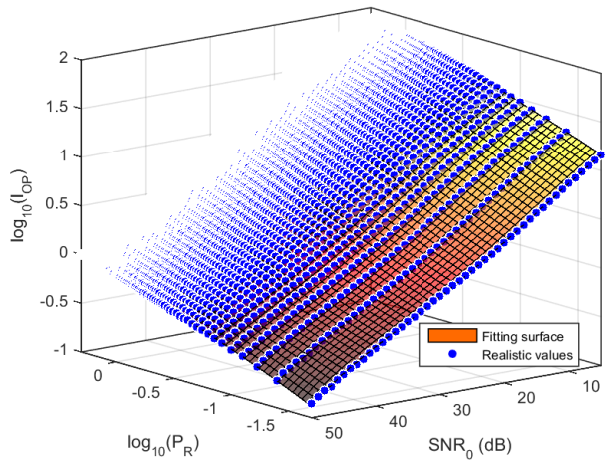}
\centering \caption{Performance illustration for the fitting of the realistic values of the open distance, where the highest degrees in (\ref{Eq:ApproximationExpression}) for $\log_{10}P_{R}$ and $\rm{SNR_0}$ are both set to be 5. The fitting surface is obtained by plotting the fitting function $\log_{10}z(x,y)$, and the realistic values represented by the dots are obtained by the exhaustive search method described in Footnote~3.} \label{Fig:poly55}
\end{figure}

\subsection{Polynomial Fitting for Open Distance}\label{Subsection:FitOfOD}
In this subsection, we attempt to obtain another expression for the open distance through least-squares approximation based on the sufficient realistic data obtained by the aforementioned exhaustive search method. From (\ref{Eq:Fitting1}), we know that the open distance is linearly dependent on the receive power and target SNR on the logarithm scale. Inspired by this fact, we utilize polynomial fitting methods to approximate the open distance over sufficient realistic values ($\log_{10} P_{R}$, $\rm{SNR_0}$, $\log_{10} l_{\text{OP}}$), expressed as
\begin{flalign}\label{Eq:ApproximationExpression}
\begin{aligned}
z(x,y) = \sum_{i=0}^{m}\sum_{j=0}^{n}f_{ij}x^{i}y^{j}, i+j \leq \max(m,n)
\end{aligned}
\end{flalign}
where $z(x,y)=\log_{10}l_{\text{OP}}$, $x=\log_{10}P_{R}$, and $y=\rm{SNR_0}$. In addition, $m$ and $n$ are the highest degrees for $x$ and $y$, respectively. However, it is worthwhile to note that (\ref{Eq:ApproximationExpression}) no longer has clear physical implications as the theoretical expression (\ref{Eq:OpenDistance}) does.

To quantitatively measure the accuracy of the above fitting, we adopt the following four widely-used GoF (Goodness of Fit) indexes \cite{Website_GoodnessOfFitting}:
\begin{itemize}
\item \textbf{SSE (Sum of Squares due to Error)}. SSE measures the total deviation of the approximation results from realistic values, with a value closer to 0 indicating a better fitting and vice versa.
\item \textbf{$R^{2}$ (Coefficient of Determination)}. It is the square of the correlation between the approximation results and realistic values, which takes values between 0 and 1, with a value closer to 1 indicating a better fitting.
\item \textbf{adj-$R^{2}$ (degree-of-freedom adjusted $R^{2}$)}. It is similar to $R^{2}$ but adjusts $R^{2}$ based on the residual degrees of freedom, which takes values less than or equal to 1, with a value closer to 1 indicating a better fitting.
\item \textbf{RMSE (Root Mean Squared Error)}. RMSE denotes an estimation of the standard deviation of the random component in the data. As with SSE, a RMSE value closer to 0 indicates a better fitting.
\end{itemize}

\begin{figure*}
\begin{equation}\label{Eq:pij}
F =
\begin{pmatrix}
1.965&-0.02957&0.0001562&-3.943\cdot10^{-5}&9.809\cdot10^{-7}&-7.009\cdot10^{-9}\\
0.2578&0.001888&0.001013&-3.835\cdot10^{-5}&3.698\cdot10^{-7}&/ \\
-0.07478&-0.002509&0.0003212&-4.849\cdot10^{-6}&/ &/ \\
-0.01852&-0.0002022&1.783\cdot10^{-5}&/&/&/\\
-0.004086&6.729\cdot10^{-5}&/&/&/&/\\
-0.0002688&/&/&/&/&/
\end{pmatrix}
\end{equation}
\end{figure*}

Table~\ref{Table:FittingResults} lists the fitting results with the highest degrees from 1 to 5. From the table, the fitting results get better when the highest degrees of $\log_{10}P_{R}$ and $\rm{SNR_0}$ grow, but it is unnecessary to set extraordinary large values for them as the increment of $m$ or $n$ will in turn increase the cost to calculate (\ref{Eq:ApproximationExpression}). Observe from Fig.~\ref{Fig:poly55} that, the approximation surface almost perfectly matches the realistic values when the degrees for $\log_{10}P_{R}$ and $\rm{SNR_0}$ come to 5. More specifically, both the SSE and RMSE are less than 0.03, and the $R^{2}$ and adj-$R^{2}$ are both equal to 1, all of which quantitatively indicate an excellent fitting performance. As a result, it suffices to take $m=5$ and $n=5$  for precisely fitting the open distance (\ref{Eq:OpenDistance}) by (\ref{Eq:pij}). The corresponding fitting coefficients, denoted by $F = \left( f_{ij} \right) $, are given by the matrix in (\ref{Eq:pij}) (note that the indexes of its row and column start from zero).

\section{Conclusions}\label{Sec:conclusion}
Unlike terrestrial radio communications, effective bandwidth of an underwater acoustic channel is dependent on the transmission range, which implies that transmission performance possibly can be improved by deploying relays along the transmission link. Following this insight, we have investigated two fundamental problems, namely when should a relay be introduced and where should it be deployed if necessary in terms of the energy and delay performance, in scenarios of direct and linear two-hop relay transmission. Regarding these two problems, we have first accurately approximated the dependence of effective bandwidth and required transmit power on the distance to formulate an energy consumption minimization problem. By the differential analysis, we have then discovered the existence of the open distance and derived a closed-form and easy-to-calculate expression for it to solve our formulated problem. Most importantly, we have strictly proved that a relay should not be introduced if the transmission distance is less than the open distance, while a relay should be deployed at the middle point of the link once the transmission distance exceeds the open distance. Although this paper considers the case of deploying one relay as the first step, our derived results also shed light on the construction of energy-efficient and delay-friendly multi-hop networks. Extensive simulation results have verified our theoretical results and the obtained data has exhibited that properly deploying a relay can dramatically reduce the network energy consumption with a negligible cost in the end-to-end delay. Moreover, we have further adopted a polynomial fitting method to precisely approximate the open distance based on the sufficient realistic data for future potential applications.

\bibliographystyle{IEEEtran}
\bibliography{IEEEabrv,Draft_Reference}
\end{document}